Crystalline Ni nanoparticles as the origin of ferromagnetism in Ni implanted ZnO crystals

Shengqiang Zhou, K. Potzger, Gufei Zhang, F. Eichhorn, W. Skorupa, M. Helm and J. Fassbender

Institute of Ion Beam Physics and Materials Research, Forschungszentrum Rossendorf, P.O. Box 510119, 01314 Dresden, Germany

Abstract:

We report the structural and magnetic properties of ZnO single crystals implanted at 623 K with up to 10 at. % of Ni. As revealed by X-ray diffraction, crystalline fcc-Ni nanoparticles were formed inside ZnO. The magnetic behavior (magnetization with field reversal and with different temperature protocol) of all samples is well explained by a magnetic Ni-nanoparticle system. Although the formation of Ni:ZnO based diluted magnetic semiconductor cannot be ruled out, the major contribution to the magnetic properties stems from crystalline nanoparticles synthesized under these implantation conditions.

## I. INTRODUCTION

Diluted magnetic semiconductors (DMS) have recently attracted huge research attention because of their potential application for spintronics devices [1, 2]. In DMS materials, transition or rare earth metal ions are substituted onto cation sites and are coupled with free carriers to yield ferromagnetism via indirect interaction [3, 4]. ZnO based DMS is particular interesting since the room temperature ferromagnetism was reported for various dopants [3-7]. However, the interpretation of the observed ferromagnetism is still controversial [8-11]. A secondary phase (metal or compound nanoparticles) induced by magnetic doping can be responsible for the measured ferromagnetism as reported in Mn:ZnO [12], Co:ZnO[13-15], and Co: $TiO_2$ [16]. Recently ferromagnetism was reported in Ni doped ZnO [5, 17-19] and $TiO_2$ [20], where however a careful characterization of structural (e.g. high sensitive method) and magnetic characteristics (e.g. temperature dependent magnetization) was only partly performed. As a contrast, Cho *et al*. questioned the origin of the ferromagnetism in Ni-doped $TiO_2$ by compositional analysis and transport measurements [21], while Zhu *et al*. reported Ni nanoparticle formation in rutile $TiO_2$ [22]. In the present paper, we attempt to clarify the ferromagnetism in Ni implanted ZnO. X-ray diffraction (XRD) and zero field cooled/field cooled magnetization (ZFCM/FCM) were used to investigate the structural and magnetic properties.

## II. EXPERIMENT

Hydrothermally grown ZnO single crystals from Crystec, Berlin were implanted with Ni ions at 623 K and with different fluences of $0.8\times10^{16}$, $4\times10^{16}$ and $8\times10^{16}$ cm$^{-2}$, respectively. The implantation energy of 180 keV yielded a projected range of $R_P$=80±34 nm, and a maximum atomic concentration of 1%, 5%, and 10%, respectively (TRIM code [23]). Virgin and implanted samples were investigated using Rutherford backscattering/channeling spectrometry (RBS/C), XRD, and SQUID (Quantum Design MPMS) magnetometry. By SQUID, the virgin ZnO is found to be purely diamagnetic with a susceptibility of $-1.48\times10^{-6}$ emu/Oe·cm$^3$. That background has been subtracted in the following discussion. XRD was performed with a Siemens D5000 diffractometer using CuK$_\alpha$ radiation. The monochromator was removed to obtain higher x-ray intensity, which is required to detect a small amount of secondary phases. The RBS spectra were collected with a collimated 1.7 MeV He$^+$ beam at a Van de Graaff accelerator with a surface barrier detector at 170°.

## III. RESULTS AND DISCUSSION

RBS/C is used to check the lattice damage after implantation. Fig. 1 shows RBS/C spectra for different fluences. The arrow labeled Zn indicates the energy for backscattering from surface Zn atoms. The implanted Ni ions cannot be detected for the very low fluence ($0.8\times10^{16}$ cm$^{-2}$, not shown), however are more pronounced as a hump in the random spectrum for a higher fluence of $4\times10^{16}$ cm$^{-2}$ and $8\times10^{16}$ cm$^{-2}$ (not shown). The channeling spectra were collected by aligning the sample to make the impinging He$^+$ beam parallel with ZnO<0001> axis. $\chi_{min}$ is the channeling minimum yield in RBS/C, which is the ratio of the backscattering yield at channeling condition to that for a random beam incidence [24]. Therefore, the $\chi_{min}$ labels the lattice disordering degree upon implantation, i.e. an amorphous sample shows a $\chi_{min}$ of 100 %, while a perfect single crystal corresponds to a $\chi_{min}$ of 1-2 %. The humps in the channeling spectra mainly come from the lattice disordering due to implantation. The $\chi_{min}$ (see Table I) increases with increasing fluence, and quantitatively confirms the applied fluence. RBS/C measurement also revealed that the ZnO host material still partly remained in a crystalline state after irradiation by Ni ions up to a fluence of $8\times10^{16}$ cm$^{-2}$ ($\chi_{min}$ of 69%).

By XRD, crystalline fcc-Ni nanoparticles were detected. Fig. 2 shows the XRD patterns

(focused on Ni(111) peak) for all samples. At a low fluence ($0.8 \times 10^{16}$ cm$^{-2}$), no evident crystalline Ni nanoparticles could be detected, while from a fluence of $4 \times 10^{16}$ cm$^{-2}$ the Ni(111) peak appeared and grew with the fluence. The inset shows a wide range scan for the highest fluence sample ($8 \times 10^{16}$ cm$^{-2}$). Ni nanoparticles were epitaxially embedded into the ZnO matrix with an orientation of Ni(111)//ZnO(0001), and no NiO particle was detected. The full width at half maximum (FWHM) of the Ni(111) peak decreased with fluence, indicating a growing of the average diameter for these nanoparticles (table I). The crystallite size is calculated using the Scherrer formula [25].

Ferromagnetism was observed in all samples. Fig. 3a shows the magnetization versus field reversal (M-H) of all samples measured at 10 K. Hysteretic behavior was observed for all three samples. The saturation moment and the coercivity increase with increasing fluence (table I). Saturation behavior is also observed at 300 K for the sample with the highest fluence (Fig. 3a inset). However neither coercivity nor remance can be observed at 300 K. The temperature dependence of the coercivity and the remanence is shown in Fig. 3b. Both values decrease drastically with increasing temperature. This is a strong indication for the superparamagnetism of a magnetic nanoparticle system. Knowing the formation of fcc-Ni from XRD, it is reasonable to assume that fcc-Ni nanoparticles are responsible for the magnetic behavior. For bulk Ni crystal (Curie temperature of ~630 K), the magnetic moment is 0.6 $\mu_B$/Ni at 0 K. If assuming the same value for Ni nanoparticles, around 27% of implanted Ni ions are in metallic state using the saturation moment at 10 K for the fluence of $4 \times 10^{16}$ cm$^{-2}$ (Table I). This fraction is much higher than that of Fe implanted ZnO, where only 13% of implanted Fe ions formed bcc-Fe nanoparticles at the same fluence and implantation energy [11].

In a solid matrix, the magnetic behavior for a single particle follows the Néel process [26, 27]. At high temperature, the system shows superparamagnetic behavior (Fig.3), while at low temperature, the system displays a slow relaxation, which can be confirmed by ZFCM/FCM measurement using a small field. Fig. 4 shows the ZFCM/FCM curves in a 50 Oe field for all samples. In order to obtain these curves, the sample was cooled in zero field from above room temperature to 5 K. Then a 50 Oe field was applied, the ZFC curve was measured with increasing temperature from 5 to 300 K, after which the FC curve was measured in the same field from 300 to 5 K with decreasing the temperature. After subtracting the diamagnetic background from the substrate, a distinct difference in ZFC/FC curves was observed. ZFC curves show a gradual increase (deblocking) at low temperature, and reach a maximum at a temperature of $T_{max}$, while FC curves continue to increase with decreasing temperature. At a much higher temperature than $T_{max}$, the FC curves still depart from corresponding ZFC curves, which distinguish the Ni particle system from a conventional spin-glass system where the FC curve merges together with the ZFC curve just at $T_{max}$ and show a plateau below $T_{max}$ [28]. The ZFC/FC curves are general characteristics of magnetic nanoparticle systems [26-28]. For a dc magnetization measurement in a small magnetic field by SQUID, the blocking temperature is given by $T_B(V) \sim K_{eff}V/k_B$, where $K_{eff}$ is the anisotropy energy density, V is the particle volume, $k_B$ is the Boltzmann constant. The $T_{max}$ is a measure of the blocking temperature of the nanoparticle system [27], and increases with the fluence, i.e. the size of nanoparticles (Table I). Indeed, it is rather difficult to calculate $T_B$ for a practical magnetic nanoparticle system. Due to size effects $K_{eff}$ is different from bulk crystals, and depends on the size of nanoparticles [26]. An inevitable size distribution of nanoparticles consequently gives rise to a distribution of $K_{eff}$ and $T_B$ [27, 29]. The dipolar interaction between nanoparticles can enhance the blocking temperature, and even induce hysteretic behavior up to room temperature [30].

## IV. CONCLUSION

To summarize, ferromagnetism was observed in Ni-implanted ZnO crystals. However, the magnetic behavior is well explained in the frame of a magnetic nanoparticle system. Crystalline fcc-Ni nanoparticles were detected by XRD. Although the formation of Ni:ZnO based DMS by ion implantation at 623 K cannot be completely ruled out, the main contribution to the ferromagnetic properties arises from these small Ni nanoparticles. This is, however, in a sharp contrast with other ZnO based DMS reports [5-7]. We have demonstrated that combining structural analysis and ZFCM/FCM measurement is a reliable approach to clarify the origin of ferromagnetism in transition metal doped ZnO.

Ref.

Table I Structural and magnetic properties for Ni-implanted ZnO with different Ni fluence. Metallic Ni fraction corresponds to the percentage of crystalline Ni compared with all implanted Ni.

| Fluence (cm$^{-2}$) | $\chi_{min}$ (RBS/C) | Crystallite size | Saturation moment at 10 K | Metallic Ni fraction | Coercivity at 10 K | $T_{max}$ (ZFCM measured) |
|---|---|---|---|---|---|---|
| 0.8x10$^{16}$ | 45% | - | 0.05 µ$_B$/Ni | 8% | 10 Oe | $\leq$ 5 K |
| 4x10$^{16}$ | 57% | 5.8 nm | 0.16 µ$_B$/Ni | 27% | 30 Oe | 16 K |
| 8x10$^{16}$ | 69% | 8.0 nm | 0.22 µ$_B$/Ni | 37% | 120 Oe | 44 K |

Fig captions

Fig. 1. RBS random (ran.) and channeling (ch.) spectra for different samples (Ni fluence is indicated).

Fig. 2 XRD patterns of Ni(111) in Ni implanted ZnO crystals with different fluence. Inset shows the wide range XRD pattern for the highest fluence sample where Ni(111) parallel with ZnO(0002) is detected and no NiO is detectable, and those sharp peaks are from ZnO due to CuK$\alpha_2$ or Tungsten emission.

Fig. 3. (a) M-H curves measured at 10 K for all samples with different Ni fluences. Inset shows the M-H curve measured at 300 K for the highest fluence sample. (b) Temperature dependent coercivity and remanence for the highest fluence sample.

Fig. 4 ZFCM/FCM curves at 50 Oe for different fluence sample. Solid symbols are ZFCM curves, while open symbols are FCM curves. Inset shows a zoom of the low temperature part of the ZFCM/FCM for the fluence of $0.8 \times 10^{16}$ cm$^{-2}$, which reveals the similar behaviour as higher fluence sample, but with a $T_{max} \leq 5$ K.

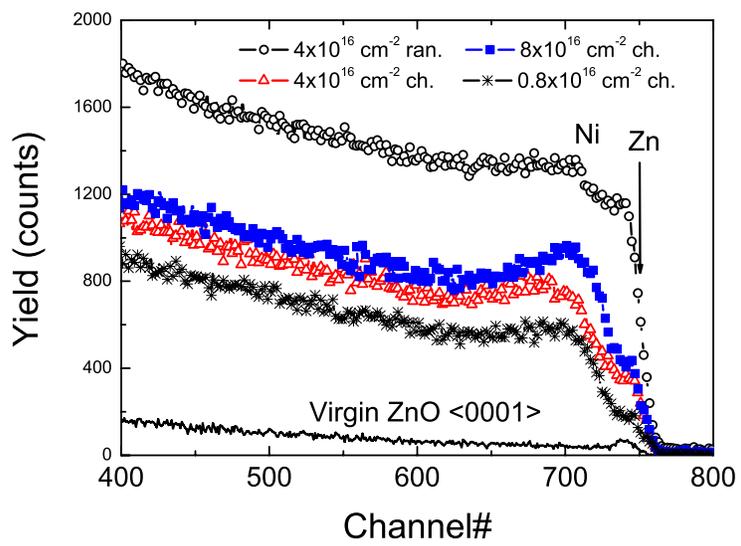

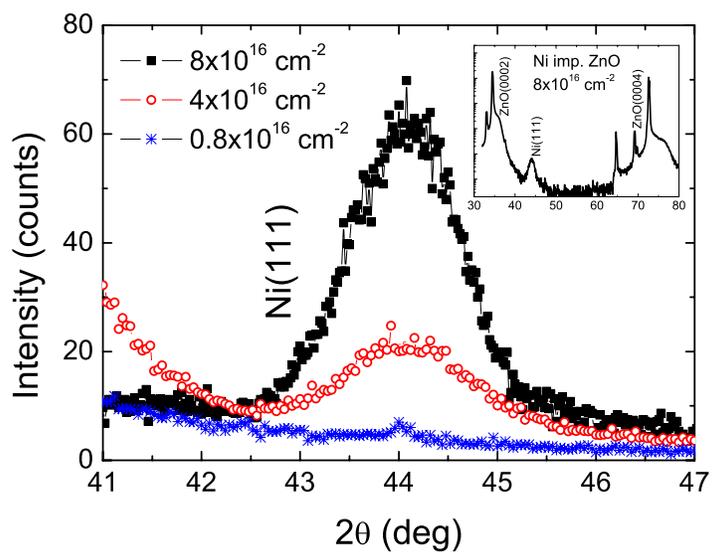

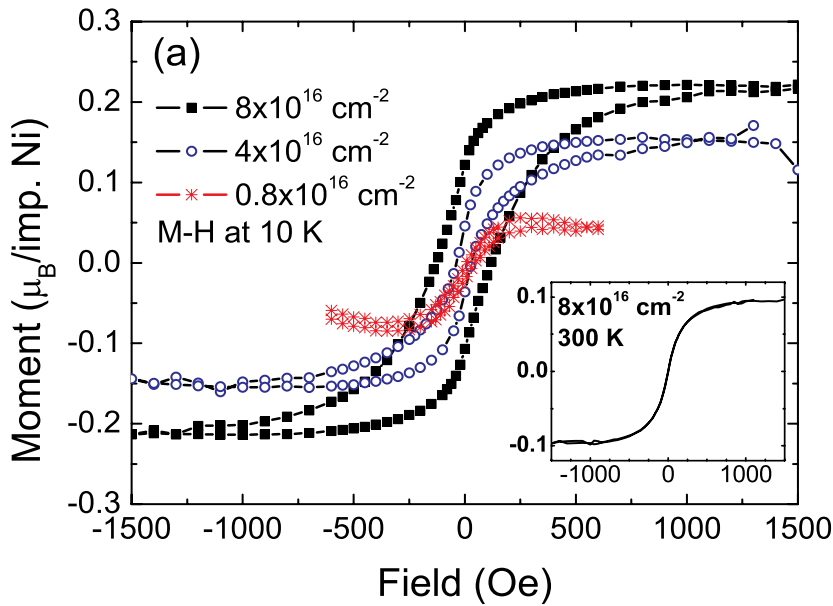

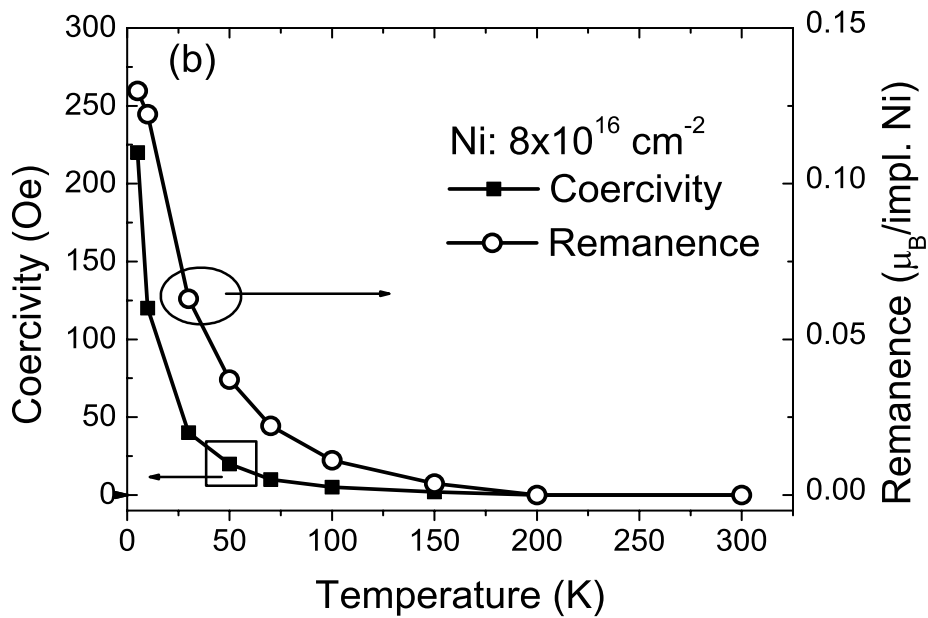

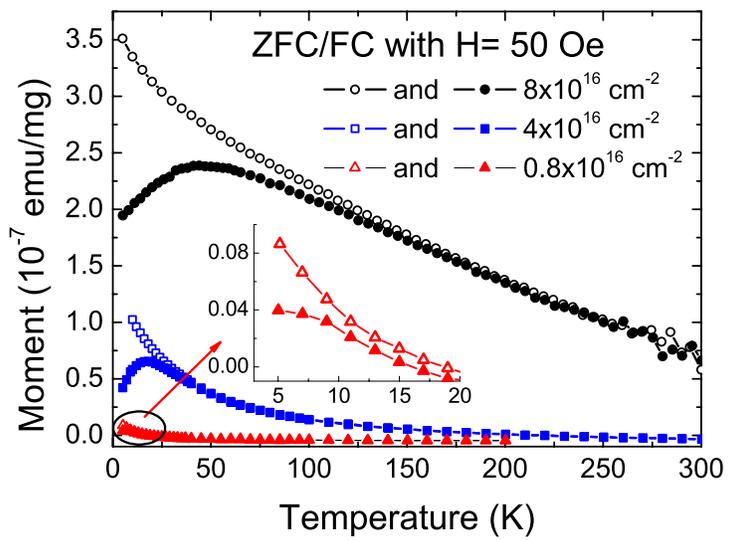